\documentclass[a4paper,11pt]{article}
\pdfoutput=1 
\usepackage{jheppub}
\usepackage{ucs}
\usepackage[utf8x]{inputenc}
\usepackage{textcomp}
\usepackage[T1]{fontenc}
\usepackage{mathtools,harmony,upgreek}
\usepackage{color,accents,stmaryrd}
\usepackage{mathrsfs}
\usepackage{amssymb}
\usepackage{amsmath}
\usepackage{dsfont}
\usepackage{indentfirst}
\usepackage{latexsym}

\newcommand{\cl}{C \kern -0.1em \ell}

\newcommand{\g}{\gamma}
\newcommand{\rf}{\lrcorner}
\newcommand{\be}{\begin{equation}}
\newcommand{\ee}{\end{equation}}
\newcommand{\ba}{\begin{array}}
\newcommand{\ea}{\end{array}}
\newcommand{\f}{\frac}

\newcommand{\beq}{\begin{eqnarray}}
\newcommand{\eeq}{\end{eqnarray}}

\newcommand{\ci}{\,\Ganz\,}

\newcommand{\w}{\wedge}

\newcommand{\G}{\Gamma}

\renewcommand{\b}{\beta}

\renewcommand{\a}{\alpha}

\definecolor{clearblue}{rgb}{0,0.5,0.9}
\definecolor{orange}{rgb}{1,0.5,0}

\title{\boldmath Spinor Fields Classification in Arbitrary Dimensions and   New Classes of Spinor
Fields on 7-Manifolds}

\author[a]{L. Bonora}
\author[b]{K. P. S. de Brito}
\author[a,c]{Rold\~ao da Rocha}

\affiliation[a]{International School for Advanced Studies (SISSA), Via Bonomea 265,
34136 Trieste, Italy}
\affiliation[b]{   CCNH, Universidade
Federal do ABC 09210-580, Santo Andr\'e, SP, Brazil}
 \affiliation[c]{CMCC, Universidade
Federal do ABC 09210-580, Santo Andr\'e, SP, Brazil}
\emailAdd{bonora@sissa.it}
\emailAdd{kelvyn.paterson@ufabc.edu.br}
\emailAdd{roldao.rocha@ufabc.edu.br}
\abstract{ A  classification of spinor fields according to  the associated bilinear covariants
is constructed in arbitrary dimensions and metric
signatures, generalizing Lounesto's 4D spinor field classification. 
In such a generalized classification a basic role is played by the geometric Fierz 
identities. In 4D Minkowski spacetime the standard bilinear covariants 
can be either null or non-null -- with the exception of the  current density which is
invariably different from zero for physical reasons --
and sweep all types of spinor fields, including Dirac, Weyl, Majorana and
more generally flagpoles, flag-dipoles and dipole spinor fields. 
To obtain an analogous classification in higher dimensions we use
the Fierz identities, which constrain the covariant bilinears in the spinor fields
and force some of them to vanish. A generalized graded Fierz aggregate is moreover
obtained in such a context simply from the completeness relation. We analyze the
particular and important case of Riemannian 7-manifolds, where
the Majorana  spinor fields turn out to have a quite special place. 
In particular, at variance with spinor fields in 4D Minkowski spacetime that
are classified in six disjoint classes,  spinors in Riemannian 7-manifolds 
are shown to be classified, according to the bilinear covariants: (a) in just
one class, in the real case of Majorana spinors; (b) in four classes,  in the
most general case. Much like new classes of spinor fields in 4D Minkowski
spacetime have been evincing new possibilities in physics, we think these new classes of
spinor fields in seven dimensions are, in particular,  potential candidates for
new solutions in the compactification of supergravity on a seven-dimensional 
manifold and its exotic versions.}
\begin{document}

\maketitle
\flushbottom

\section{Introduction}


Generalizing Fierz identities in 4D Minkowski spacetime has led to new
interesting results on spinor fields, with respect to the textbook ones,  and unexpected applications likewise. It is therefore
natural to try to do the same in higher dimensions. Fierz identities for form-valued spinor bilinears were considered in arbitrary 
dimensions and metric signatures \cite{1,Babalic:2013fm}, using the geometric algebra, being also based in the developments \cite{okubo11,rand}.
The most general Fierz identities were further used to construct independent
effective four-fermion interactions that contain  
spin-3/2 chiral fields \cite{3}. Besides, the existence of
two natural bilinear forms on the space of spinors were shown in  \cite{5} to be
 related with elements of the exterior algebra. All the Fierz identities can be
reduced to a single equation using the extended Cartan map \cite{10}.
Reciprocally, spinors are reconstructed from Fierz identities \cite{14} 
and quantum tomography for Dirac spinors were considered  in this context as well
\cite{mosna}.
General Fierz identities were further used to find completeness and
orthogonality relations   \cite{6,7}, where an equivalence
between spinor and tensor representation of various quantities have been also
constructed.
Moreover, Fierz identities are useful to calculate scattering 
amplitudes \cite{8}, being thus employed to calculate electroweak interactions.
The inverse problem can be  solved  likewise by using the Fierz identities
\cite{9}. Various Fierz identities are further needed to show the invariance 
of the $D=11$, $N=1$ supergravity action under a local supersymmetry
transformation, and the supercovariance of the
fermion field equation \cite{11}. In supersymmetric gauge theories,
supersymmetry constraints imply the existence of certain Fierz identities for
real Clifford 
algebras \cite{12}. These identities hold merely for 2, 4, 8 and 16 supercharges
\cite{Top}. Fierz identities were also generalized for non-integer dimensions in
\cite{13}. 

The prominent relevance of Fierz identities can be moreover measured by their
role on the recent emergence of new kinds of spinor fields. 
The subject concerning those new spinor fields and their applications 
has been widening, mainly since the middle of the last decade.  
Fierz identities were  used by Lounesto to classify spinor fields in
Minkowski spacetime according to the bilinear covariants \cite{lou2}.
Indeed, Lounesto showed that spinor fields can be accommodated in  six disjoint
classes, that encompass all the spinor fields in Minkowski spacetime. The first
three types of spinor fields in such classification are named Dirac spinor
fields: this is actually a generalization, as it does not restrict to the standard Dirac
spinor field, which is an eigenspinor of the parity operator, providing hence further physical solutions for the Dirac equation. Indeed such three
classes of regular spinor fields appear as solutions of the Dirac equation in different contexts. They
are characterized by either the scalar or the pseudoscalar
(or even both) bilinear covariants being nonzero. The other three classes of
(singular) spinors are known as flag-dipole, flagpole and dipole spinor fields,
and have both the scalar and pseudoscalar bilinear covariants vanishing. The latter
classes contain, besides a rich geometric structure, spinor fields with new
dynamics. Flagpole spinor fields have been recently considered in cosmology
\cite{shank}, being  explored as candidates for dark matter in various contexts
\cite{alex,exotic,lee2,lee1}, wherein Elko and Majorana spinor fields evince
prominent roles  \cite{daRocha:2005ti}. Flag-dipole ones are typified for
instance by recently found new solutions of the Dirac
equation in ESK gravities \cite{esk}. Dipole spinor fields include Weyl spinor
fields as their most known representative.  All the  spinor classes have been
lately thoroughly characterized \cite{Cavalcanti:2014wia}. A complete overview
of this classification with further applications in field theory and
gravitation can be found in \cite{daSilva:2012wp}, being also further explored
in the context of black hole thermodynamics \cite{bht}. Indeed, black hole
tunnelling methods were studied for Elko spinor fields as special type of
flagpoles \cite{bht}, which play an essential role in constructing various theories
of gravity naturally arising from supergravity \cite{daRocha:2009gb,
daRocha:2007sd}. Flagpoles spinor fields and Lounesto spinor field
classification are moreover discussed in the context of the  instanton Hopf
fibration \cite{hopf}, and experimental signatures of the type-5 spinors in such
a classification are related to the Higgs boson at LHC \cite{m1}. An up-to-date
overview on a special class of such spinor fields can be found in
\cite{crevice} and references therein.

Fierz identities make it possible to deduce a classification of spinors on the
spinor bundle associated to manifolds of arbitrary dimensions, departing from the case
proposed by Lounesto, that holds solely for Minkowski spacetime \cite{lou2}. 
The aim of the present paper is two-fold: besides generalizing Lounesto's spinor field
classification for spacetimes of arbitrary dimension and metric signatures and
proposing a graded  Fierz aggregate, we focus in particular on the noteworthy 
case of spinor fields on seven-dimensional manifolds, in particular those
on the 7-sphere. This is motivated by their wide physical applications, for instance, 
 in $D=11$  supergravity \cite{SUGRA,SUSYBr,engl}. In fact, spontaneous
compactifications of $D=11$ supergravity \cite{SUGRA} on Riemannian
7-manifolds are well-known to contain all the degrees of freedom of the massless sector of gauged
$N = 8$ supergravity theory \cite{engl}. Explicit forms
of the Killing spinors can be also obtained  in the supergravity theory that
 admits an AdS$_4\times S^7$ solutions.
$S^7$ spinors and the Ka$\check{\rm c}$-Moody algebra of $S^7$ were also
considered on the parallelizable 7-sphere \cite{ced,eu1}.

We use the Fierz identities constraining bilinear covariants
 constructed by  spinor fields in arbitrary  dimensions \cite{1},  that force some of
the respective bilinear covariants to be zero and study the important case of 
seven dimensions, in particular classifying Majorana 
 spinor fields. 
 New kinds of classes of spinor fields are hence obtained, which are
hidden if one considers only the real spin bundle on Riemannian 7-manifolds.

This paper is organized as follows: in Section II, the bilinear covariants are
used to revisit the classification of spinor fields in Minkowski spacetime, 
according  to the Lounesto's classification prescription, and the Fierz
aggregate and its related boomerang are defined. In Section III, the bilinear
covariants associated to spinor fields in arbitrary dimensions and metric
signatures are introduced and all the possibilities for them are listed. In
Section IV, the geometric Fierz identities are employed and from the admissible
pairings between spinor fields the number of classes in the spinor field
classification are constrained. We study the case of Majorana spinor fields on
Riemannian 7-manifolds and define the graded Fierz aggregate as a particular
case of the completeness condition. We conclude that Majorana spinors in seven dimensions pertain to solely one class, according the bilinear
covariants, as some of these bilinears are identically zero. Spinor fields  on
these 7-manifolds can be classified in four classes, departing from the classification for the real
spin bundle. One of the new classes encompasses the Majorana spinor fields and the
others provide new candidates  for physical solutions, for instance, in supergravity. 

\section{Bilinear Covariants in Minkowski Space-time}

In order to fix the notation, consider an oriented manifold $(M,g)$,
where the metric $g$ has signature $(p,q)$, and its associated tangent
[cotangent] bundle $
TM$ [$T^*M$], having sections consisting of $n$-dimensional $(n=p+q)$ real vector spaces. Denoting sections of the exterior bundle by
$\sec\bigwedge (TM)$, given a $k$-vector $a \in \sec\bigwedge^k(TM)$,  the grade
involution is defined by $\hat{a}=(-1)^{k}a$ and the {reversion}  by
$\tilde{a}=(-1)^{[[k/2]]}a$, where [[$k$]] stands for the integral part of $k$. The
 conjugation is the composition of the two previous morphisms. 
Moreover, when ${g}$ is extended from $\sec\bigwedge^1(TM)=\sec T^*M$ to
$\sec\bigwedge(TM)$, and by considering  $a ,b ,c \in \sec \bigwedge(V)$, the
{left [right] contraction} can be  defined by ${g}(a \lrcorner b,c)={g}(b
,\tilde{a}\wedge c )\; [{g}(a \llcorner b ,c )={g}(b ,a \wedge \tilde{c})$].  
The Clifford product between a vector field $ v \in \sec\bigwedge^1(TM)$ and a multivector  
$a \in \sec\bigwedge(TM)$ is prescribed  by $ v \circ a = v \wedge a+ v 
\lrcorner a $. The dual Hodge operator $\star:\sec\bigwedge(TM)
\rightarrow\sec\bigwedge(TM)$ is defined by 
$a\w\star b = g(a,b)$. The Grassmann algebra $(\bigwedge (TM),{g})$ endowed
with the Clifford product is denoted by $\cl_{p,q}$, the Clifford algebra
associated with $\sec\bigwedge^1(TM)\simeq \mathbb{R}^{p,q}$.

When the Minkowski spacetime is considered, the set $\{{e}_{\mu }\}$ represents sections of the frame bundle
$\mathbf{P}_{\mathrm{SO}_{1,3}^{e}}(M)$ and  $\{\theta ^{\mu }\}$ is the dual 
basis $\{{e}_{\mu }\}$, namely, $\theta ^{\mu }({e}_{\mu })=\delta^\mu_{\;\nu}$.
 Classical spinor fields are objects in the carrier space associated to a
$\uprho={\rm D}^{(1/2,0)}\oplus {\rm D}^{(0,1/2)}$ representation of the Lorentz group, and can be thought as being sections of the vector bundle
$\mathbf{P}_{\mathrm{Spin}_{1,3}^{e}}(M)\times _{\uprho }\mathbb{C}^{4}$. Moreover, the classical spinor fields carrying either the
${\rm D}^{(0,1/2)}$ or the ${\rm D}^{(1/2,0)}$ representation of the Lorentz group are
sections of the vector bundle $\mathbf{P}_{\mathrm{Spin}_{1,3}^{e}}(M)\times
_{\uprho^{\prime }}\mathbb{C}^{2},$ where 
$\uprho ^{\prime }$ stands for either the ${\rm D}^{(1/2,0)}$ or the ${\rm D}^{(0,1/2)}$
representation of the Lorentz group. Given a
spinor field 
$\psi \in \sec \mathbf{P}_{\mathrm{Spin}_{1,3}^{e}}(M)\times_{\uprho
}\mathbb{C}^{4}$, the bilinear covariants  are the  following sections of the
exterior algebra bundle $\bigwedge(TM)$ \cite{lou2,moro,cra}:
\begin{subequations}
\begin{eqnarray}
\sigma &=& \bar{\psi}\psi\,,\label{sigma}\\
\mathbf{J}&=&J_{\mu }\theta ^{\mu }=\bar{\psi}\gamma _{\mu }\psi\, \theta
^{\mu}\,,\label{J}\\
\mathbf{S}&=&S_{\mu \nu }\theta ^{\mu}\wedge\theta^{ \nu }=\tfrac{1}{2}i\bar{\psi}\gamma _{\mu
\nu }\psi \,\theta ^{\mu }\wedge \theta ^{\nu }\,,\label{S}\\
\mathbf{K}&=& K_{\mu }\theta ^{\mu }=i\bar{\psi}\gamma_{5}\gamma _{\mu }\psi
\,\theta ^{\mu }\,,\label{K}\\\omega&=&-\bar{\psi}\gamma_{5}\psi\,,  \label{fierz}
\end{eqnarray}\end{subequations}
where $\bar\psi=\psi^\dagger\gamma_0$,
$\gamma_5:=\gamma_0\gamma_1\gamma_2\gamma_3$ and the set $\{\mathbf{1},\gamma
_{\mu },\gamma _{\mu }\gamma _{\nu },\gamma _{\mu }\gamma _{\nu}\gamma _{\rho
},\gamma _{5}\}$ ($\mu <\nu <\rho $) is a basis for
${M}(4,\mathbb{C})$ satisfying  $\gamma_{\mu }\gamma _{\nu
}+\gamma _{\nu }\gamma_{\mu }=2\eta_{\mu \nu }\mathbf{1}$ and the Clifford
product is denoted here by juxtaposition \cite{moro}.

The
space-like 1-form $\mathbf{K}$ designates the spin direction,  the
2-form $\mathbf{S}$ denotes the well-known intrinsic angular momentum, and the time-like 1-form $\mathbf{J}$ stands for the current of probability. The bilinear
covariants  satisfy the Fierz identities \cite{cra, lou2} 
\begin{equation}\label{fifi}
-(\omega+\sigma\gamma_{5})\mathbf{S}=\mathbf{J}\wedge\mathbf{K},\qquad\mathbf{K}^{2}+\mathbf{J}^{2}
=0=\mathbf{J}\llcorner\mathbf{K},\qquad
\mathbf{J}^{2}=\omega^{2}+\sigma^{2}\,.  
\end{equation}
\noindent When $\omega=0=\sigma$, a spinor field is said to be singular, and
regular otherwise. 

Lounesto \cite{lou2} classified spinor fields into six disjoint classes. In the
classes (1), (2), and (3) beneath it is implicit that $\mathbf{J},$ $\mathbf{K}$
and $\mathbf{S}$ are simultaneously different from zero, and in the classes (4), (5), and (6) just $\mathbf{J}\neq 0$:

\begin{itemize}
\item[1)] $\omega\neq0,\;\;\;  \sigma\neq0$\qquad\qquad\qquad\qquad\qquad4) $
\omega=\sigma=0 , \;\;\;\mathbf{K}\neq 0, \;\;\;\mathbf{S}\neq0$%
\label{Elko11}
\item[2)] $\omega = 0,\;\;\;
\sigma\neq0$\label{dirac1}\qquad\qquad\qquad\qquad\qquad5) $\omega=\sigma=0 ,
\;\;\;\mathbf{K}=0,\;\;\; \mathbf{S}\neq0$%
\label{tipo41}
\item[3)] $\omega \neq0, \;\;\;\sigma= 0$\label{dirac21}
\qquad\qquad\qquad\qquad\qquad\!6) $\omega=\sigma=0 , \;\;\; \mathbf{S}=0,
\;\;\; \mathbf{K} \neq 0$%
\end{itemize}
\noindent 
Spinor fields of types-1, -2, and -3 are called Dirac spinor fields  whilst
spinor fields of types-4, -5, and -6 are flag-dipoles, flagpoles and dipole
spinor fields, respectively \cite{lou2}. It is worthwhile to emphasize that the
naming ``Dirac spinors'' in Lounesto's classification is wider than the one
adopted in textbooks, where Dirac spinors are eigenstates of the parity
operator. The first physical example of flag-dipole spinor fields has been found very recently in
\cite{esk} as solutions of the Dirac equation in a $f(R)$ background with
torsion. Moreover, Majorana and Elko spinor fields reside in the class of
spinors of type-5 \cite{daRocha:2005ti}, and Weyl spinor fields are a particular
case of a type-6 dipole spinor fields \cite{lou2}, wherein  further spinor fields have been scarcely scrutinized. It is  also worthwhile to point out
that in four and in six dimensions pure spinors coincide with Weyl spinors due
to an accident in these dimensions \cite{lou2,hopf}, while there are 
quadratic constraints that pure spinors obey in higher dimensions \cite{budinich}. In
particular, the constraints in ten dimensions  play an important role in
Berkovits' approach to superstrings \cite{bk1,bk2}.

A multivector field \cite{lou2}   
\begin{equation}
Z=\omega\gamma_{5}+i\mathbf{K}\gamma_{5}+i\mathbf{S}+\mathbf{J}+\sigma\label{boomf}
\end{equation} is called a Fierz aggregate 
when $\omega, \mathbf{S}, \mathbf{K}, \mathbf{J}, \sigma$  fulfil the Fierz
identities (\ref{fifi}).  Additionally, if $\gamma^{0}Z\gamma^{0}=Z^{\dagger}$, 
the Fierz aggregate is named a boomerang \cite{lou2}. When singular spinor
fields  are scrutinized, the Fierz identities are replaced by the most assorted
expressions \cite{cra}: 
\beq\nonumber
Z\gamma_{\mu}Z=4J_{\mu}Z,\qquad Z^{2} =4\sigma Z,\qquad iZ\gamma_{\mu\nu
}Z=4S_{\mu\nu}Z,\nonumber\\ 
-Z\gamma_{5}Z=4\omega Z,\qquad  iZ\gamma_{5}\gamma_{\mu}Z=4K_{\mu}Z.
\label{boom}
\eeq

\section{Bilinear Forms with Spinors in Arbitrary Dimensions and Metric Signatures}

Going to arbitrary dimensions, one starts from the spin bundle ${\it S}$ associated to a
manifold $(M,g)$. A crucial role is played by the K\"ahler-Atiyah bundle
($\sec\bigwedge(TM),\circ$),  where the Clifford product
shall be denoted by $\circ$. 
The spin bundle {\it S} is defined upon the even
K\"ahler-Atiyah bundle ($\mathring{\bigwedge}(TM),\circ$) and has module structure specified
by a morphism $\gamma:(\bigwedge(TM),\circ)\rightarrow ({\rm End}(S),\,\Ganz\,)$
  (for more details, see, e. g., \cite{1,Babalic:2013fm,bonora,moro,face,lazaroiu}).
In addition, a direct sum decomposition $S=S_0\oplus S_1$ is provided by
an idempotent endomorphism R $\in \Gamma({\rm End}(S))$ (here $\Gamma({\rm End}(S))$
denotes the space of smooth sections of ${\rm End}(S)$) -- which for some dimensions and
signatures is usually  identified with the volume element $ {\gamma^{n+1}}$  \cite{1}. The sub-bundles $S_0$ and $S_1$ are determined by the 
eigenvalues $\pm1$ of R, and the above direct sum decomposition is said to be non-trivial if both $S_0$ and $S_1$ 
are different from zero. This is equivalent to saying that the Clifford algebras
$\mathcal{C}\ell_{p,q}$ constructed on the  cotangent bundle on each point of
$M$ are universal \cite{bt}.  The restriction $\mathring{\gamma}:
\sec\mathring{\bigwedge}(TM)\rightarrow {\rm End}(S)$ 
is  named spin endomorphism if it commutes with $\gamma(\xi)$, for all
$\xi\in\sec \mathring{\bigwedge}(TM)$ \cite{1}. 

Spin projectors are defined by $
\Pi_\pm= \frac{1}{2}( I  \pm  {\rm R})$, where $I$ denotes
the identity operator on $S$, providing the direct sum  $ 
S=S^+\oplus S^-$, where $
S^\pm= \Pi_\pm (S)$. 
The sections of $S^\pm$ are called
 [symplectic]  {Majorana-Weyl spinors}  when $p-q\equiv 0\mod 8$ [$p-q\equiv
4\mod 8$], while  the sections of $S^+$ are known as  [symplectic] Majorana
spinors when  
 $p-q\equiv 7\mod 8$ [$p-q\equiv 6\mod 8$]. 
Classical spinors $S_{p,q}$ of the Clifford bundle of $M$  are well-known to be
elements of the irreducible representation space of the component of the group Spin$(p,q)$ connected to the identity, and their
classification 
can be summarized as follows:\medbreak 
\begin{center}
\begin{tabular}{||c||c|c|c|c||} \hline \hline 
$\begin{matrix} p-q \\ \text{mod} \, 8 \end{matrix}$ & 0 & 1 & 2 
& 3 \\ \hline 
\fbox{$ \begin{matrix}   S_{p,q}   \end{matrix}$}  & 
$\begin{matrix} \mathbb{R}^{2^{[(n-1)/2]}} \\ \oplus \\
\mathbb{R}^{2^{[(n-1)/2]}} \end{matrix}$& 
$\mathbb{R}^{2^{[(n-1)/2]}}$& 
$\mathbb{C}^{2^{[(n-1)/2]}}$& 
$\mathbb{H}^{2^{[(n-1)/2]-1}}$ \\ \hline 
$\begin{matrix}p-q \\ \text{mod} \, 8\end{matrix}$ & 4 
& 5 & 6 & 7 \\ \hline 
 \fbox{$ \begin{matrix}   S_{p,q}   \end{matrix}$}   & 
$\begin{matrix} \mathbb{H}^{2^{[(n-1)/2]-1}} \\ \oplus \\
\mathbb{H}^{2^{[(n-1)/2]-1}}\end{matrix}$ & 
$\mathbb{H}^{2^{[(n-1)/2]-1}}$& 
$\mathbb{C}^{2^{[(n-1)/2]}}$ & 
$\mathbb{R}^{2^{[(n-1)/2]}}$ \\ \hline\hline 
\end{tabular}\\
\medskip
{\small Table I. {Classical Spinors Classification Table -- Real Case} 
($p + q = n$)}
\end{center}
\medbreak
The ring of quaternions is denoted by $\mathbb{H}$.  
 For the complex case the classification is well-known to be simpler:\medbreak
 \begin{center}
\label{eq.4.40.1}
\begin{tabular}{||c|c||} \hline \hline  \fbox{$ \begin{matrix}   n = 2k  
\end{matrix}$}  & 
$\mathbb{C}^{2^{k-1}} \oplus \mathbb{C}^{2^{k-1}}$\\ \hline
 \fbox{$ \begin{matrix}   n = 2k +1  \end{matrix}$ }   & 
$\mathbb{C}^{2^k}$
\\ \hline\hline 
\end{tabular}\\
\medskip
{\small Table II. {Classical Spinors Classification Table -- Complex Case} }
\end{center}\bigskip

Now consider  the complex structure $J\in \Gamma({\rm End}(S))$, which in
particular is given by $
J=\pm \gamma^{n+1}=\pm \gamma^1\,\Ganz\, \cdots \,\Ganz\, \gamma^n$ when
$p-q\equiv3, 7\mod 8$ \cite{1}, and  an endomorphism $D$ on the spin bundle
that satisfies for all $\xi\in \sec\bigwedge (TM)$ the following expressions
\cite{1,okubo}
\beq
&& D\,\Ganz\,D=(-1)^{\frac{1+p-q}{4}} I,\qquad\quad D\,\Ganz\,
\gamma(\xi)=\gamma(\hat\xi)\,\Ganz\, D\,.\label{dddd}
\eeq\noindent Such expressions are taken into account
hereupon, being essential to define both the classification of spinor fields and
the graded Fierz aggregate as well.

Starting from an orthonormal local coframe $\{e^a\}_{a =1}^{ n }\subset
\mathbf{P}_{\mathrm{SO}_{p,q}^{e}}(M)$, recall from \cite{alek1} that a
non-degenerate bilinear pairing $ B$ on the spin bundle $S$ is named 
admissible if the following requirements hold:  
a)  $ B$ is either symmetric or skew-symmetric; b)  if $p-q\equiv 0,4,6,7\mod
8$,  then $S^+$ and $S^-$ are either isotropic or orthogonal with respect to $B$
  \cite{1};  
c) for any $\xi\in \sec\bigwedge (TM)$ one has the transpose relation 
$
\gamma(\xi)^\intercal =\gamma(\underaccent{\tilde}{\xi})$ if and only if
$ 
 B(\gamma(\xi)\psi,\psi')=  B(\psi,\gamma(\underaccent{\tilde}{\xi})
\psi')$, 
where $\underaccent{\tilde}{\xi}$ stands for  either the usual reversion
$\tilde{\xi}$, if $B$ is symmetric, or the Clifford conjugation $\bar\xi$
otherwise. 

A more general pairing can be taken into account, by complexifying its
restriction to the real bundle $S^+$. 
In fact, by adopting hereon the notation $\psi,\psi'\in \Gamma(S),$   where
$\Gamma(S)$ denotes the space of smooth sections of the spin bundle $S$, the
bilinear pairing $\b_0$ on $S$ is obtained \cite{1}
\beq
\!\!\!\!\!\!\beta_0(\psi,\psi')\!=\! B\!\left(\!{}^{\rm (Re)}\psi,\!{}^{\rm (Re)}\psi'\right)\!-\!
B\!\left(\!{}^{\rm (Im)}\psi,\!{}^{\rm (Im)}\psi'\right)\!+\!i\!\left[B\!\left(\!{}^{\rm
(Re)}\psi,\!{}^{\rm (Im)}\psi'\right)\!+\! B\!\left(\!{}^{\rm (Im)}\psi,\!{}^{\rm
(Re)}\psi'\right)\right],\label{formaa}
\eeq\noindent  where ${}^{\rm (Re)}\psi=\f{1}{2}(\psi+D(\psi))$ and ${}^{\rm
(Im)}\psi=\f{1}{2}(\psi-D(\psi))$ are the real and 
imaginary parts of $\psi$, respectively \cite{1}.

 For an ordered set of indexes $(\alpha_1,\ldots,\alpha_k)$, $1\leq \alpha_k\leq n=p+q$, the
notation 
$e^{\alpha_1\ldots \alpha_k}= e^{\alpha_1}\wedge\cdots\wedge e^{\alpha_k}$ and $\gamma^{\alpha_1\ldots
\alpha_k}=  \gamma^{\alpha_1}\,\Ganz\,\cdots \,\Ganz\, \gamma^{\alpha_k}$ shall be adopted
hereupon, with analogous expressions for their respective contravariant counterparts.
The product ``$\Ganz$'' may be also denoted by juxtaposition, being
used  explicitly solely when we want to emphasize the underlying structure.

In the previous section the usual spinor conjugate
$\bar\psi=\psi^\dagger\gamma^0$ represents the spinor dual to $\psi$. In Minkowski spacetime 
it has this form in order to produce a Lorentz invariant
quantity. In arbitrary dimensions it can have a much more general form, which plays
a prominent role in the framework of
bilinear forms. In fact, a spin-invariant product can always be written as \cite{bt}
$$\beta(\psi,\psi')=a^{-1}\tilde{\psi}\psi'= \psi^\dagger a^{-1}\psi',$$  
where $\psi, \psi'$ are spinor fields and $a\in\G({\rm End}(S))$.  Since
$\tilde{a}=a$ and $\mathring{\mathring{b}}=ab^\dagger a^{-1}$, where
$\;\mathring{\mathring{}}\;$ indicates an arbitrary adjoint involution, 
then $a^\dagger=a$ \cite{bt}. In this way, the spinor conjugation can be written
more generally as $
\bar\psi=a^{-1}\tilde{\psi}=\psi^\dagger a^{-1}$, and we can use this definition
to write the most general bilinear  on $S$: 
\beq\label{formab}
\beta_k(\psi,\psi^\prime)=B(\psi,\gamma_{\alpha_1\dots\alpha_k} \psi^\prime)=
{\bar{\psi}}{\gamma}_{\alpha_1\dots\alpha_k} {\psi^\prime}\,.
\eeq\noindent  In the well-known case described in Section II, $\psi$ is a
spinor field in Minkowski spacetime, and usually $\bar{\psi}=\psi^\dagger\g_
0$. 

On the spin bundle $S$, bilinear covariants are thus more generally defined as
follows, given  $\psi\in S$:
 \begin{subequations} \beq
 \Omega_1 &=&\bar{\psi}\psi\\
J_{\a_1}&=&\bar{\psi}\g_{\a_1}\psi\\
S_{\a_1\a_2}&=&\bar{\psi}\g_{\a_1\a_2}\psi\\
 &\vdots&\nonumber\\
 S_{\a_1\dots\a_{d+1}}&=&\bar{ \psi}\g_{\a_1\ldots\a_{d+1}} \psi\,,\qquad
d<k-1\\
 &\vdots&\nonumber\\
 S_{\a_1\ldots\a_k}&=&\bar{ \psi}\g_{\a_1\ldots\a_k}\psi \\
 K_{\a_1\ldots\a_{l-1}}&=&\bar{\psi}\g_{\a_1\ldots\a_{l-1}}\g_{n+1}\psi
\\
  &\vdots&\nonumber\\
 K_{\a_1\dots\a_m}&=&\bar{\psi} \g_{\a_1\ldots\a_m}\g_{n+1}\psi\,, \quad m<l-1\\
 &\vdots&\nonumber\\
 K_{\a_1}&=& \bar{\psi}\g_{\a_1}\g_{n+1}\psi\\
\Omega_2 &=&\bar{\psi}\gamma_{n+1}\psi\;.
 \eeq \end{subequations}

These bilinear covariants are used to defined the corresponding Fierz aggregate 
\beq
{\rm Z}=\Omega_1 +J_{\mu}\g^{\mu}+S_{\mu_1\mu_2}\g^{\mu_1\mu_2}+\dots +
S_{\mu_1\ldots\mu_p}\g^{\mu_1\ldots\mu_p}+\nonumber\\
+K_{\a_1\ldots\a_{q-1}}\g^{\a_1\ldots\a_{q-1}}\g_{{n+1}}+\dots
+K_{\mu}\g^{\mu}\g_{{n+1}}+\Omega_2 \g_{{n+1}}\,,
\eeq \noindent that reduces to the standard Fierz aggregate (\ref{boomf}) when
$n=4$. Similarly to  the case on Minkowski spacetime, 
when $\gamma^{0}{\rm Z}\gamma^{0}={\rm Z}^{\dagger}$ the above generalized Fierz aggregate will be called a generalized boomerang.  

In the next section we shall see that a graded Fierz aggregate can be evinced
exclusively from the completeness relation. In addition, most of such bilinear
covariants that define it will be shown to be null, as a consequence of
constraints imposed by the geometric Fierz identities \cite{1}. It forces the number of
spinor field classes to be reduced, being obstructed by the geometric Fierz
identities. We will see in the next section that the 4D Fierz identities for regular spinor fields -- given by (\ref{fifi}) -- and for any kind of spinor fields -- provided by (\ref{boom}) --    can be generalized  to the geometric Fierz identities in arbitrary dimensions \cite{1}, and new classes of Fermi fields on 7-manifolds can be evinced.

\section{Where are Majorana Spinor Fields in the Generalized Spinor Field
Classification? New Classes of Spinor fields}

When $p-q\equiv  7\mod 8$,
the endomorphism $D$ is a real structure that defines the complex conjugate via $D(\psi)={}^{(\rm Im)}\psi$, and can be identified to the spin endomorphism R discussed in Section 3 \cite{1}. The projectors $P_\pm = \frac{1}{2}(I\pm D)$ are hence responsible to split a spinor field in real and imaginary components given by respectively by $P_\pm(\psi)=\psi^\pm$. 
The real vector bundles $S^\pm \equiv P_\pm(S)$ are thus used to  identify the spin bundle
$S=S^+\oplus S^-$ to 
the complexification of the real bundle $S^+$ of
Majorana spinor fields \cite{1}. In particular when $n=7$, other useful pairings can be
defined from a basic admissible pairing, denoted hereon for the sake of
simplicity by $B$, as \cite{1} 
\beq\label{b1b2}
 B':=  B \,\Ganz\,( I \otimes J), \qquad B'':= - B \,\Ganz\, [ I \otimes (J\,\Ganz\, D)]\,,\qquad B{'''}:= B \,\Ganz\, ( I \otimes
D)\,\,.
\eeq\noindent We can use some of them to define the bilinears (\ref{formaa}) and
(\ref{formab}). However, might the higher dimensional analogues of Lounesto's
classes of spinor fields 
provide indeed a different spinor fields classification? 
We shall answer in an affirmative way this question and discuss later such new possibilities, showing
that at least two choices in (\ref{b1b2}) are equivalent under Hodge duality.

When one chooses $\psi$ to be a Majorana spinor, the non null bilinear pairings
can be reduced through the fact that $
  B (\psi,\g^{\alpha_1\ldots \alpha_k}\psi)=0$ except if   $k$ is even \cite{1}.
Given any admissible bilinear pairing $ B $ on $S$, the
endomorphisms $A_{\psi |\psi'}$ of the spin bundle $S$ have been defined in
\cite{1}:
\be
A_{\psi_1 |\psi_2}(\psi):= B (\psi,\psi_2)\psi_1\,,\quad \text{for all}\;\;\;\;\;
\psi,\psi_1,\psi_2 \in \Gamma(S)\,,
\ee\noindent and play an important role  to 
determine the geometric Fierz identities, encrypted in the expressions 
\begin{equation}
A_{\psi_1 |\psi_2}\,\Ganz\, A_{\psi_3 |\psi_4}= B
(\psi_3,\psi_2)A_{\psi_1 |\psi_4}\,,
\end{equation} 
as shall be briefly reviewed in the sequel \cite{1}.

Consider now the completeness relation 
\begin{equation}\nonumber
A_{\psi |\psi^\prime}=\frac{\ell}{2^n}\sum_k  \frac{1}{k!} (-1)^{k}  B 
(\psi,\gamma_{\alpha_1\ldots \alpha_k} \psi')e^{\alpha_1\ldots
\alpha_k}\,,\end{equation} where either $\ell = 2^{\llceil\frac{n}{2}\rrceil}$, if
$p-q=0,1,2$, or  $\ell = 2^{\llceil\frac{n}{2}\rrceil+1}$ otherwise, where
$\llceil\frac{n}{2}\rrceil \equiv \frac{n(n-1)}{2}  \mod\,2$. This expression
mimics the Fierz aggregate (\ref{boomf}). 
Moreover, every element in the space $\Gamma({\rm End}(S))$, in particular the
$A_{\psi |\psi'}$, can be split  uniquely as  $
A_{\psi |\psi'}=D\,\Ganz\, A^1_{\psi |\psi'}+A^0_{\psi |\psi'}$ \cite{1}, 
where
\begin{subequations}
\begin{eqnarray}
 A  ^0_{\psi |\psi'} &=&\frac{\ell}{2^n}\sum_k  \frac{(-1)^{k}}{k!}   B (\psi,
\gamma_{\alpha_1\ldots \alpha_k}\psi')e^{\alpha_1\ldots \alpha_k}\,,\label{eoo1} \\
 A  ^1_{\psi |\psi'} &=&\frac{\ell}{2^n}\sum_{k}  \frac{1}{k!}(-1)^{\left(k+\frac{1+p-q}{4}\right)} 
 B (\psi, D\,\Ganz\, \gamma_{\alpha_1\ldots
\alpha_k}\psi')e^{\alpha_1\ldots \alpha_k}\,. \label{eoo2}
\end{eqnarray}
\end{subequations}

The Fierz
identities (\ref{fifi}) -- for regular spinor fields in Minkowski spacetime -- or more generally (\ref{boom}) --- for any kind of spinor fields in Minkowski spacetime -- can be  generalized for arbitrary dimensions, being 
hence provided by \cite{1}:
\begin{subequations}\begin{eqnarray}
 \widehat{A  ^0_{\psi_1 |\psi_2}}\circ  A  ^1_{\psi_3 |\psi_4} +   A 
^1_{\psi_1 |\psi_2}\circ  A  ^0_{\psi_3 |\psi_4} 
&=&   B (\psi_3,\psi_2) A  ^1_{\psi_1 |\psi_4}\,,\label{fp1}\\
  A  ^0_{\psi_1 |\psi_2}\circ  A  ^0_{\psi_3 |\psi_4} + (-1)^{\frac{1+p-q}{4}}
\widehat{A  ^1_{\psi_1 |\psi_2}}\circ  A  ^1_{\psi_3 |\psi_4} 
&=&   B (\psi_3,\psi_2) A  ^0_{\psi_1 |\psi_4}\,.\label{fp2}
\end{eqnarray}\end{subequations}
The notation for both elements in $\sec\bigwedge^k(TM)$
\begin{subequations}\beq
 A ^{0,k}_{\psi |\psi'}&=& \frac{1}{k!} (-1)^{k}   B (\psi,
\gamma_{\alpha_1\ldots \alpha_k}\psi')e^{\alpha_1\ldots
\alpha_k}\,,\label{eok1}\\
 A ^{1,k}_{\psi |\psi'}&=& \frac{1}{k!} (-1)^{\frac{1+p-q}{4}+k}  B (\psi,
D\,\Ganz\, \gamma_{\alpha_1\ldots \alpha_k}\psi')e^{\alpha_1\ldots
\alpha_k}\,,\label{eok}  
\eeq \end{subequations}
will be employed accordingly,  
in order to make it possible to write $
 A  ^{\lambda}_{\psi |\psi'} =\frac{\ell}{2^n}\sum_k 
A^{\lambda,k}_{\psi |\psi'}$, for $\lambda = 0,1$. 

As we are interested in determining the nature of Majorana spinor fields
according to bilinear covariants in 7-manifolds, we focus in the particular, however important case \cite{ced} of $n=p+q = 7+0$. The
case of a Majorana spinor on a Riemannian 7-manifold
arises, for example, in the of $N=1$
compactifications of $M$-theory on 7-manifolds
\cite{MT, BJ, TH}, permitting a 
geometric characterization by means of the reduction  of the
structure group of
$\sec\bigwedge^1(TM)$ from the orthogonal group O(7) to the exceptional one G$_2$ \cite{1}.

 Moreover, 
the expression 
 \beq
\label{ddd}
\varphi_k:= \vert A^{0,k}_{\psi |\psi}\vert = \frac{1}{k!}B (\psi,\gamma_{\alpha_1\ldots
\alpha_k}\psi)e^{\alpha_1\ldots \alpha_k} \eeq\noindent equals zero except if
$k$ is even. Together with the symmetry of $  B $, this shows that the element $ A ^{0,k}\equiv A ^{0,k}_{\psi |\psi}$
vanishes
except if 
 $k=0,3,4,7$.  Combining this with
\eqref{ddd}, it implies that, for $\psi$ a Majorana spinor, the
forms $\varphi_k$ equal zero except when  $k=0$ or $k=4$. By regarding $\psi$ 
normalized
 such that $  B (\psi,\psi)=1$, it follows  that $ A^{0,0}=1$ and the following
bilinear can be defined \cite{1}:
\beq
\label{phi4}
\varphi_4=\frac{1}{4!}  B (\psi,\gamma_{\alpha_1\alpha_2\alpha_3
\alpha_4}\psi)e^{\alpha_1\alpha_2\alpha_3\alpha_4}\,,
\eeq
which are the components of the first generator $
 A^0=\frac{1}{16}(1+\varphi_4)$ of the so called Fierz algebra represented in (\ref{eoo1}, \ref{eoo2}) \cite{1}, where 
 $ A^\lambda\equiv A^\lambda_{\psi |\psi}$. 
Moreover, since $
 A^1= A^0$, then the Fierz identities (\ref{fp1}, \ref{fp2}) are  concomitantly
equivalent and imply 
that 
$(\varphi_4+1)\circ(\varphi_4+1)=8(\varphi_4+1)$ \cite{1}. 

Now, recall that the product $\Delta _k:\sec\bigwedge (TM)\times \sec\bigwedge
(TM)\rightarrow \sec\bigwedge (TM)$ is defined iteratively by 
\beq
\chi\,\Delta
_{k+1}\,\vartheta=\f{1}{k+1}g^{ab}(e_a\lrcorner\chi)\,
\Delta_k\,(e_b\lrcorner\vartheta)\,,\qquad\chi, \vartheta\in \sec\bigwedge (TM)\,,
\eeq\noindent where $g^{ab}$ denotes the metric  tensor coefficients. 
By fixing $\Delta _0=\wedge$ as being the exterior product,  the next terms
are for instance
 provided by \beq
\chi\,\Delta
_1\,\vartheta&=&g^{ab}(e_a\lrcorner\chi)\w(e_b\lrcorner\vartheta)\,,\nonumber\\
\chi\,\Delta _2\,\vartheta&=&
\f{1}{2}g^{ab}g^{cd}[e_a\rf(e_c\rf\chi)]\w[e_b\rf(e_d\rf\vartheta)]\,.\nonumber
\eeq
When the Clifford product is written as 
$\varphi_4\circ\varphi_4= ||\varphi_4||^2-\varphi_4\,\Delta_2\,\varphi_4$,
the Fierz identities
correspond to the following  conditions \cite{1}:
\beq
\label{erz}
\varphi_4\,\Delta_2\,\varphi_4=-6\varphi_4\,,\qquad\qquad ||\varphi_4||^2=7\,.
\eeq\noindent They are the root to establish the spinor fields classification
according to the bilinear covariants.

 In  the case here to be analyzed $n=p+q=7$, we already know that $\varphi_k=0$ except for the values $k\in\{0,3,4,7\}$.  In addition, due to the restriction 
\beq
  B (\psi,J\,\Ganz\,\gamma_{\alpha_1\ldots\alpha_k}\psi)=-  B (\psi,J\,\Ganz\,
D\,\Ganz\,\gamma_{\alpha_1\ldots\alpha_k}\psi)= 0\;\text{except it $k=2j+1$, $k\in\mathbb{Z}$}\,,
\eeq
we obtain that unless $k=3$ or $k=7$
the form ${ B}'(\psi,\g_{\alpha_1\ldots \alpha_k}\psi)$ vanishes. Hence,
alternatively we could have chosen any other of the bilinear pairings in
(\ref{b1b2}) in order to define the bilinear covariants. For instance, choosing $B'$ yields the following definition: 
\beq
 \label{phij}
\check{\varphi}_k=\f{1}{k!}  B (\psi, J\ci\g_{\alpha_1\ldots
\alpha_k}\psi)e^{\alpha_1\ldots \alpha_k}\in\sec\bigwedge^k(TM)\,.\eeq\noindent 
However, the Hodge duality $\star\xi=\tilde\xi J$, where $J=\gamma^{n+1}$,
 for our case $n=7$ implies that the alternative homogeneous forms
\beq
\check{\varphi}_3&=&\f{1}{3!}  B (\psi,
J\ci\g_{\alpha_1\alpha_2\alpha_3}\psi)e^{\alpha_1\alpha_2\alpha_3}\\
\check{\varphi}_7 &=&\f{1}{7!}  B (\psi, J\ci\g_{\alpha_1\dots
\alpha_7}\psi)e^{\alpha_1\dots \alpha_7}
\eeq do not vanish likewise. 
A similar reasoning \cite{1} implies that the other bilinear pairings in
(\ref{b1b2}) contain no information besides the ones provided by (\ref{ddd}). 
The Fierz identities take now the form 
\beq
\check{\varphi}_3\Delta
_i\varphi_4 = 0\quad (i=1,3)\,,\qquad \check{\varphi}_3\wedge\varphi_4&=&||\check{\varphi}_3||^2\gamma^{8}\,.
\eeq
Since $\check{\varphi}_3=\star\varphi_4$ implies that
$\|\check{\varphi}_3\|=\|\varphi_4\|$, Eq. (\ref{erz}) asserts that the 3-form
$\check{\varphi}_3$ is non null likewise, and a similar reasoning implies that
also $\check{\varphi}_7\neq 0$, as $\star 1 =\check{\varphi}_7$. 

Hence, only the bilinears
\begin{eqnarray}
 \varphi_0&=&  B  (\psi,\psi)\label{B00}\\
\varphi_4&=&\f{1}{4!}  B
(\psi,\g_{\alpha_1\alpha_2\alpha_3\alpha_4}\psi)e^{
\alpha_1\alpha_2\alpha_3\alpha_4}
 \label{B0}
\end{eqnarray}
are non null. Eq. (\ref{B00}) is the higher dimensional analogue 
of its Minkowski spacetime version provided by Eq.(\ref{sigma}).
Thus the Fierz identities (\ref{erz}) imply, in particular, that
$\varphi_4\neq0$ and only one type of  Majorana spinor results  according to a generalized  spinor
field classification:
\beq
\varphi_0\neq 0\in\sec\bigwedge^0(TM),\qquad\quad\label{neq2} \varphi_4\neq
0\in\sec\bigwedge^4(TM).
\eeq
In fact, such class of Majorana spinor fields according to the bilinears in the Clifford bundle ${\cal C}\ell_{7,0}$ is provided by:
\beq
\!\!\!\!\!\!\!\!\!\!\!\!\!\!\varphi_0\neq 0, \quad \varphi_1=0, \quad\varphi_2= 0, \quad\varphi_3= 0,\quad
\varphi_4\neq 0,\quad\varphi_5=0, \quad\varphi_6=0, \quad
\varphi_7=0\,,\label{class}\eeq\noindent or equivalently, if Eq.(\ref{phij}) is
taken into account,
\beq
\!\!\! \!\!\!\!\!\!\!\!\!\!\!\!\!\!\!\!\!\quad\check{\varphi}_7\neq0,  \quad\check{\varphi}_6=0,
\quad\check{\varphi}_5=0,  \quad\check{\varphi}_4=0, \quad\check{\varphi}_3\neq
0,\quad\check{\varphi}_2= 0,\quad\check{\varphi}_1=0, \quad\check{\varphi}_0 =
0\,,\label{class1}
\eeq
where the Hodge duality $\check{\varphi}_k= \star\varphi_{7-k}$ is utilized.

If we use the results from \cite{1} and given $\iota\in\mathbb{Z}$,
\beq
 B (\psi,\gamma_{\alpha_1 \ldots\alpha_k}\psi) 
=
\begin{cases}
 B\left({}^{\rm (Re)}\psi,\gamma_{\alpha_1 \ldots\alpha_k}{}^{\rm
(Im)}\psi\right)+  B\left({}^{\rm (Im)}\psi,\gamma_{\alpha_1
\ldots\alpha_k}{}^{\rm (Re)}\psi\right)\,, \,\text{if $k=2\iota$}\\
 B\left({}^{\rm (Re)}\psi,(J\;\Ganz\;\gamma_{\alpha_1 \ldots\alpha_k}){}^{\rm
(Re)}\psi\right)-  B\left({}^{\rm (Im)}\psi,(J\;\Ganz\;\gamma_{\alpha_1
\ldots\alpha_k}\right){}^{\rm (Im)}\psi)\,,\\\hspace{9cm} \,\text{if
$k=2\iota+1$}\,,\nonumber
\end{cases}
 \eeq
 and
\beq
 B(\psi,J\,\Ganz\,\gamma_{\alpha_1 \ldots\alpha_k}\psi) 
=\begin{cases}
- B\left({}^{\rm (Re)}\psi,\gamma_{\alpha_1 \ldots\alpha_k}{}^{\rm
(Im)}\psi\right)+  B\left({}^{\rm (Im)}\psi,\gamma_{\alpha_1
\ldots\alpha_k}{}^{\rm (Re)}\psi\right)\,, \,\text{if $k=2\iota$}\\
 B\left({}^{\rm (Re)}\psi,(J\;\Ganz\;\gamma_{\alpha_1 \ldots\alpha_k}){}^{\rm
(Re)}\psi\right)+  B\left({}^{\rm (Im)}\psi,(J\;\Ganz\;\gamma_{\alpha_1
\ldots\alpha_k}){}^{\rm (Im)}\psi\right)\,, \\\hspace{8.5cm}\,\text{if $k=2\iota+1$}\,,
\end{cases}\label{fgh}\nonumber
 \eeq
 Eq. (\ref{formab}) that describes the bilinear covariant coefficient of degree $k$ can be
thus generalized, in order to 
 encompass the complex case, providing the higher degree 
 generalization of (\ref{formaa}): 
 \beq\label{formac}
\upbeta_k(\psi,\gamma_{\alpha_1\ldots\alpha_k}\psi')&=& B\left({}^{\rm
(Re)}\psi,\gamma_{\alpha_1\ldots\alpha_k}{}^{\rm (Re)}\psi'\right)-
B\left({}^{\rm (Im)}\psi,\gamma_{\alpha_1\ldots\alpha_k}{}^{\rm
(Im)}\psi'\right)\\&&\nonumber\qquad\qquad+i\left[B\left({}^{\rm
(Re)}\psi,\gamma_{\alpha_1\ldots\alpha_k}{}^{\rm (Im)}\psi'\right)+
B\left({}^{\rm (Im)}\psi,\gamma_{\alpha_1\ldots\alpha_k}{}^{\rm
(Re)}\psi'\right)\right]\,.\eeq\noindent 
 By using the above  results, the bilinear covariants 
 can be now extended from the standard Majorana spinor fields in $\psi\in
\Gamma(S^+)$ to sections of
$\Gamma(S)$, by identifying now 
\beq
\upvarphi_k:=
\frac{1}{k!}\upbeta_k(\psi,\gamma_{\alpha_1\ldots\alpha_k}\psi)e^{
\alpha_1\ldots\alpha_k}\,.\eeq As both terms in the real part and also both
terms in the imaginary part in (\ref{formac}) as well can cancel each other, in the complex version it is possible that the bilinears $\varphi_0$ and $\varphi_4$ can now be different from zero. Hence,  four
classes  of spinor fields $\psi\in\Gamma(S)$ are found, concerning the
classification  in Riemannian 7-manifolds   using the
constraints above, as it is showed below:
\begin{subequations}
\beq & \upvarphi_0=0,\quad\upvarphi_4=0,\label{c11}\\
& \upvarphi_0=0,\quad\upvarphi_4\neq0,\label{c12}\\
& \upvarphi_0\neq0,\quad\upvarphi_4=0,\label{c13}\\
& \upvarphi_0\neq0,\quad\upvarphi_4\neq0\,.\label{c14}\eeq
\end{subequations} It is implicit in (\ref{c11}-\ref{c14}) that all other $\upvarphi_k=0$ for $k=1,2,3,5,6,7$. Moreover, the spinor field classification according to the bilinears
$\check{\upvarphi}_k$, defined by substituting Eq.(\ref{phij}) in (\ref{formac}), is identical to the one provided by the $\upvarphi_k$.
The above class (\ref{c14}) encompasses the sole spinor
field class (\ref{class}), and reduces to it when we restrict the field $\psi$ to
be an element of the  bundle $\Gamma(S^+)$, namely, a Majorana spinor field.

Since $  B (\psi, \g_{\alpha_1\ldots \alpha_k}\psi)$ vanishes except when 
$k\in\{0,3,4,7\}$, the graded Fierz aggregate, that is defined  by
\beq\label{boomg}{\rm Z}=\f{\ell}{2^{n}}\sum_{k}(-1)^{k} B 
(\psi,\g_{\alpha_1\ldots \alpha_k}\psi)e^{\alpha_1\ldots \alpha_k},\eeq where
the sum is ordered in $k$, has clearly the terms $ B  (\psi,\g_{\alpha_1\ldots
\alpha_k}\psi) = 0$ for such values of $k$. The above expression coincides with
its  Minkowski spacetime version (\ref{boomf}) provided by Lounesto \cite{lou2}.

After classifying the spinor fields in Riemannian 7-manifolds, they can be used
for defining one Lagrangian on $S^7$ for matter fields. According to \cite{Top},
terms in a Lagrangian defined in this way depend on which realization is taken
for the matter spinor fields. 
The classification of spinor fields  in 7-manifolds  can be very useful in order to
study the behaviour of fields in AdS$_4\times S^7$ or, more generally, in AdS$_4\times M^7$, where $M^7$ denotes a 7-manifold.  

\section{Conclusion}

Spinor fields on a manifold $(M,g)$, with arbitrary dimension and arbitrary
metric signature have been classified according to the bilinear covariants. It
encompasses the celebrated Lounesto's spinor field classification for Minkowski
spacetime, generalizing it  to arbitrary dimensions and metric signatures. The
geometric Fierz identities \cite{1} limit the amount of classes of spinor
fields in such a generalized classification, which is explicitly analyzed for
the important case of Majorana spinor fields on Riemannian 7-manifolds.  A
generalized graded Fierz aggregate is also obtained in such a context simply
from the completeness relation, and we analyze the particular and prominent case
of  7D. In this case, the higher the spacetime  dimension, the
lesser the number of classes of spinor fields. 

Despite the generalizations regarding Fierz identities were known, 
analysis of the spinor fields themselves had been lacking up to the middle of the
last decade. Since then new models had been proposed, as for
instance a candidate for the field theory of some mass dimension  one fermions. A complete
characterization and identification of these new spinor fields as elements of the
Lounesto's classes \cite{daRocha:2005ti} have introduced 
in the literature new possibilities, further investigated in e. g.
\cite{alex,lee2,daRocha:2005ti,exotic,esk,Cavalcanti:2014wia,daSilva:2012wp,bht,
daRocha:2007sd}. In fact, spinor fields in the same class can present
completely different dynamics. For instance, Elko spinor fields pertain to the class 5 in
Lounesto's classification, and satisfy a coupled system of Dirac-like equations,
whereas Majorana spinor fields are also spinor fields in the class 5, but
satisfy the well-known Majorana equations.  Recently the first physical example of a
flag-dipole particle, which is a type-4 spinor field in the Lounesto's
classification, has been found as the solution of the Dirac equation with
torsion in a $f(R)$ background \cite{esk}. Type-6 spinor fields encompass for
instance Weyl spinor fields, but the complete dynamics of all classes is nevertheless 
undetermined. The first important step toward a complete characterization of all
possible dynamics of spinor fields in Minkowski spacetime  has been accomplished
by explicitly obtaining the most general type of spinor fields in each class of
Lounesto's classification \cite{Cavalcanti:2014wia}. 

Thus, the study of the Lounesto's spinor fields classification has opened a huge
path to discover unexpected 
new physical features and to propose candidates for new particles in Minkowski spacetime  
\cite{alex,lee2,lee1,daRocha:2005ti,exotic,esk,Cavalcanti:2014wia,daSilva:2012wp,daRocha:2007sd,m1}. With this motivation we have proposed a much more general
classification as well, encompassing pseudo-Riemannian spacetimes of arbitrary
dimensions and metric signatures. In particular, as the subject of $S^7$ spinor
fields is very rich \cite{ced,SUGRA}, we investigated where are the Majorana
spinor fields in such a classification according to the bilinear covariants, and
we concluded that the geometric Fierz identities obstruct the existence of more
than one precise class, determined by (\ref{class}), asserting that for instance
that spinor fields studied in \cite{ced,SUGRA,SUSYBr} reside in such class. In
these papers, the 3-form bilinear is the torsion tensor that works as a gauge
potential. In the most general case that we analyzed, by taking spinor fields in
the spin bundle over $M$, more   three types of spinor fields with potential new
properties are achieved. 
As the singular spinor fields in Lounesto's classification
were studied in  exotic structures \cite{alex,exotic}, it is natural to relate the
new classes of spinor fields in 7-manifolds derived in (\ref{c11}-\ref{c14})  to
their exotic version \cite{Yamagishi:1983dw}; however this is beyond of the scope of 
the present paper.

\acknowledgments

R. da Rocha is grateful for the CNPq grant No. 303027/2012-6 and No. 473326/2013-2 which
has provided partial support, to CAPES for the \emph{Bolsa Processo N$^{o}$}
10942/13-0, and to SISSA for the hospitality. K. P. S. de Brito acknowledges the UFABC and CAPES grants.

\end{document}